\documentclass[aps,showpacs,11pt]{revtex4}
\usepackage{dcolumn}
\usepackage{graphicx}
\usepackage{amsmath}
\usepackage{amsfonts}
\usepackage{amssymb}
\usepackage{psfrag}
\usepackage{wrapfig}
\usepackage{subfigure}
\usepackage{makeidx}
\usepackage{bm}
\usepackage{epsf}
\usepackage{hyperref}
\usepackage{color}
\usepackage{multirow,dcolumn}
\usepackage{graphicx}

\begin{document}

\title{Microscopic explanation for black hole phase transitions via Ruppeiner geometry: two competing factors--the temperature and repulsive interaction among BH molecules}

\author{Yong Chen, Haitang Li and Shao-Jun Zhang}
\email{2111709024@zjut.edu.cn; lhaitang@zjut.edu.cn; sjzhang84@hotmail.com}
\affiliation{Institute for Theoretical Physics $\&$ Cosmology, Zhejiang University of Technology, Hangzhou 310023, China}

\date{\today}

\begin{abstract}

Charged dilatonic black hole (BH) has rather rich phase diagrams which may contain zeroth-order, first-order as well as reentrant phase transitions (RPTs) depending on the value of the coupling constant $\alpha$ between the electromagnetic field and the dilaton. We try to give a microscopic explanation for these phase transitions by adopting Ruppeiner's approach. By studying the behaviors of the Ruppeiner invariant $R$ along the co-existing lines, we find that the various phase transitions may be qualitatively well explained as a result of two competing factors: the first one is the low-temperature effect which tends to shrink the BH and the second one is the repulsive interaction between the BH molecules which, on the contrary, tends to expand the BH. In the standard phase transition without RPT, as temperature is lowered, the first kind of factor dominates over the second one, so that large black hole (LBH) tends to shrink and thus transits to small black hole (SBH); While in the RPT, after the LBH-SBH transition, as temperature is further decreased, the strength of the second factor increases quickly and finally becomes strong enough to dominate over the first factor, so that SBH tends to expand to release the high repulsion and thus transits back to LBH. Moreover, by comparing the behavior of $R$ versus the temperature $T$ with fixed pressure to that of ordinary two-dimensional thermodynamical systems but with fixed specific volume, it is interesting to see that SBH behaves like a Fermionic gas system in cases with RPT, while it behaves oppositely to an anyon system in cases without RPT. And in all cases, LBH behaves like a nearly ideal gas system.

\end{abstract}


\maketitle

\section{Introduction}

In the past decades, black hole (BH) thermodynamics has become a hot topic in modern theoretical physics, as its deep connection to quantum gravity as well as other areas of fields via AdS/CFT~\cite{Maldacena:1997re,Gubser:1998bc,Witten:1998qj}, such as condensed matter physics~\cite{Hartnoll:2009sz,Herzog:2009xv,McGreevy:2009xe,Horowitz:2010gk,Cai:2015cya}, quantum information theory~\cite{Swingle:2009bg,Swingle:2012wq,Qi:2013caa}, QCD~\cite{Mateos:2007ay,Gubser:2009md,CasalderreySolana:2011us}, cosmology~\cite{Banks:2004eb} and etc. Classically, BH is a dark spacetime region enclosed by the so-called event horizon which is a one-way surface so that particles, even with velocity of light, while pass through the surface into the region can never return anymore. So, it is like a black body without any radiation. Remarkably, while taking into account the quantum effect, Hawking found that BH radiates like a thermal object with temperature proportional to its surface gravity~\cite{Hawking:1974sw}. Previous to Hawking's seminal work, Bekenstein had proposed that BH should have entropy proportional to its horizon area to accommodate the second law of thermodynamics~\cite{Bekenstein:1973ur}. Further studies reveal that not only it does have temperature and entropy, it also obeys four mechanics resembling the four laws of thermodynamics~\cite{Bardeen:1973gs}. Moreover, as ordinary thermodynamical systems, BHs may also possess rich phase structures and phase transitions between in. In Ref.~\cite{Hawking:1982dh}, Hawking and Page found that there is a first-order phase transition between thermal radiation and Schwarzschild BH in AdS spacetime, now refers to Hawking-Page transition. Interestingly, it is interpreted as confinment-deconfinement transition in the dual quantum field system in AdS/CFT~\cite{Witten:1998zw}. Since these pioneering work, amounts of efforts have been devoted to study thermodynamics and phase transitions of various BHs in diverse gravity theories.

The phase diagram of BH becomes even much more richer when it is studied in the so-called extended phase space, where the negative cosmological constant is considered as the thermodynamical pressure. Within this framework, BHs behave in many ways analogous to a variety of chemical phenomena of ordinary thermodynamical systems, such as solid-liquid phase transition, Van der Waals fluid, triple points, reentrant phase transition (RPT), heat engines and etc. And thus BH thermodynamics in this framework now sometimes is referred to as BH chemistry~\cite{Kastor:2009wy,Kubiznak:2014zwa,Mann:2016trh}. In the extended phase space, the Hawking-Page transition is similar to that of solid-liquid transition, with the thermal radiation and Schwarzschild-AdS BH respectively playing roles of solid and liquid states~\cite{Kubiznak:2014zwa}. When one consider charged AdS BH, it is found that there is a first-order large-black-hole/small-black-hole (LBH-SBH) transition in a canonical ensemble, which is in many ways analogous to the liguid-gas transition of Van der Waals fluid~\cite{Chamblin:1999tk,Chamblin:1999hg,Dolan:2011xt,Kubiznak:2012wp}. In Refs.~\cite{Gunasekaran:2012dq,Altamirano:2013ane}, it is found that at certain range of temperature, a LBH-SBH-LBH phase transition appears which is similar to the RPT observed in ordinary thermodynamical systems, with the LBH-SBH transition being first-order and SBH-LBH transition zeroth-order. There have been amounts of work on discussing various phase transitions for diverse BHs, for more details please refer to the review Ref.~\cite{Kubiznak:2016qmn} and refs therein.

For an ordinary thermodynamical system, its thermodynamical properties, such as the entropy and phase transition, at least in principle, can be explained statistically from its microscopic structure. However, for BH, its underlying microscopic structure is still a mystery. There are some attempts trying to explain its thermodynamical properties and phase transitions microscopically. For example, the entropy of certain extremal BH can be explained microscopically in string theory~\cite{Strominger:1996sh}. However, for more general BHs, it is still an open question. And it is believed that a full understanding of microscopic structure of the BH requires a complete quantum gravity theory. However, recently there are some attempts trying to understand phase transitions of BHs microscopically via an alternative way~\cite{Mansoori:2013pna,Mansoori:2014oia,Wei:2015iwa,Mansoori:2016jer,
Zangeneh:2016snh,Dehyadegari:2016nkd,Zangeneh:2016fhy,Miao:2017cyt,Miao:2017fqg,Miao:2018qyh,Hendi:2015xya}, namely the Ruppeiner approach~\cite{Ruppeiner:1979,Ruppeiner:1995zz}. In Ruppeiner approach, one can use the Ruppeiner invariant $R$ (the Ricci scalar of Ruppeiner geometry) to explore the interactions between microscopic constituents. The sign of $R$ indicates the dominant interaction: $R>0$, $R<0$ and $R=0$ respectively imply repulsion, attraction and no interaction. And the magnitude of $R$ measures the average number of Planck areas on the horizon that are correlated. For more information on its applications on BH thermodynamics, please refer to the review Ref.~\cite{Ruppeiner:2013yca}. In Ref.~\cite{KordZangeneh:2017lgs}, by comparing the behaviors of $R$ versus the temperature $T$ for Born-Infeld-AdS BHs and singly-spinning Kerr-AdS BHs, the authors found that two properties are responsible for the appearance of RPTs. One is that, in cases with RPT, $R<0$ for the SBH near the transition line meaning attractive interaction between the possible BH molecules, while in cases without RPT, $R>0$ meaning repulsive interaction. The other one is that for the range of pressure where RPT occurs, $R$ for SBH will be much more negative than for the other range of pressure where no RPT occurs. Moreover, in cases with RPT, the SBH behaves like a Bosonic gas while in the case without RPT, it behaves like a quantum anyon system. This study gives us some insights into the possible microscopic origin of BH RPTs. Then it is natural and interesting to apply this approach on other BHs with RPTs to see if this method can still be valid.

In this work, we will apply the Ruppeiner approach to study the phase transitions in charged dilatonic BHs~\cite{Sheykhi:2007wg,Sheykhi:2007gw}. We will see that the charged dilatonic BH has rather rich phase structures depending on the value of the coupling constant $\alpha$ between the dilaton and the electromagnetic field. As $\alpha$ is increased from zero to the upper bound, several types of phase diagram appear consequently with/without zeroth-order, first-order phase transitions or RPT. Different from the standard definition of the Ruppeiner metric, which is defined on the $(M, Q)$-space as done in Ref.~\cite{KordZangeneh:2017lgs} (see also related work), we define the metric instead on the $(M, P)$-space as $Q$ is fixed in our cases. By calculating the Ruppeiner invariant $R$ along the co-existing lines, we find that the various phase transitions may be qualitatively well explained as a result of two competing factors: the first one is the low-temperature effect which tends to freeze the BH molecules and thus shrink the BH and the second one is the repulsive interaction between the BH molecules which, on the contrary, tends to expand the BH. In the standard phase transition without RPT, as temperature is lowered, the first kind of factor dominates over the second one, so that LBH tends to shrink and thus transits to SBH; While in the RPT, after the LBH-SBH transition, as temperature is further decreased, the strength of the second factor increases quickly and finally becomes strong enough to dominate over the first factor, so that SBH tends to expand to release the high repulsion and thus transits back to LBH. Moreover, by studying the behaviors of $R$ versus the temperature, we find that SBH behaves like a Fermi gas system in cases with RPT, while it behaves oppositely to an anyon system in cases without RPT. And in all cases, LBH behaves like a weakly interacting Fermi gas system.

This work is organized as follows. In Sec.II, charged dilatonic BHs in Einstein-Born-Infeld-dilaton theory and Einstein-Maxwell-dilaton theory we considered in this work will be reviewed. Then its phase structures and various phase transitions will be investigated in Sec.III. In Sec.IV, by adopting the Ruppeiner approach, we try to give some microscopic explanations for the various phase transitions. The last section is devoted to summary and discussions.

\section{Review of phase transitions of dilatonic black holes}

In this section, we will review the phase transitions in charged dilatonic BHs which have been studied thoroughly in Refs.~\cite{Dehyadegari:2017flm,Momennia:2017hsc,Dayyani:2017fuz}. We consider the following four-dimensional Einstein-Born-Infeld-dilaton action~\cite{Sheykhi:2007gw,Sheykhi:2006dz}
\begin{eqnarray}
S = \frac{1}{16\pi G} \int d^4 x \sqrt{-g} \left[R - 2 (\partial \Phi)^2 - V(\Phi) + 4 \beta^2 e^{2 \alpha \Phi} \left(1 - \sqrt{1 + \frac{e^{-4 \alpha \Phi F^2}}{2 \beta^2}}\right)\right],
\end{eqnarray}
where the dilaton field $\Phi$ is coupled to the electromagnetic field $A_\mu$ with coupling constant $\alpha$. $\beta$ is the so-called Born-Infeld parameter. In the limit $\beta \rightarrow \infty$, the action reduces to the Einstein-Maxwell-dilaton action~\cite{Chan:1995fr}. The dilaton potential takes the form
\begin{eqnarray}
V(\Phi) = 2 \Lambda e^{2\alpha \Phi} + \frac{2 \alpha^2}{b^2 (\alpha^2 -1)} e^{2\Phi /\alpha},
\end{eqnarray}
with $b$ being an arbitrary constant and $\Lambda$ playing the role of cosmological constant. With this specific dilaton potential, the equations of motion allow the following spherical symmetric BH solution
\begin{eqnarray}
ds^2 &=& -f(r) dt^2 + f^{-1} (r) dr^2 + r^2 R^2(r) (d\theta^2 + \sin^2\theta d\phi^2),\nonumber\\
A_\mu &=& (A_t(r),0,0,0),\qquad \Phi = \Phi(r).
\end{eqnarray}
with
\begin{eqnarray}
f(r) &=& - \frac{\alpha^2 + 1}{\alpha^2 -1} \left(\frac{b}{r}\right)^{-2\gamma} - \frac{m}{r^{1-2\gamma}} + \frac{(\alpha^2 +1)^2 r^2}{\alpha^2 -3} \left(\frac{b}{r}\right)^{2\gamma} \left\{\Lambda + 2\beta^2 \left[_2{\cal F}_1 \left(-\frac{1}{2}, \frac{\alpha^2 -3}{4}, \frac{\alpha^2 +1}{4}, -\eta\right) -1\right]\right\},\nonumber\\
R(r) &=& e^{\alpha \Phi},\nonumber\\
A_t (r) &=& \frac{q}{r}\ _2{\cal F}_1 \left(\frac{1}{2}, \frac{\alpha^2 + 1}{4}, \frac{\alpha^2 +5}{4}, -\eta\right),\nonumber\\
\Phi (r) &=& \frac{\gamma}{\alpha} \ln\left(\frac{b}{r}\right).
\end{eqnarray}
Here $_2{\cal F}_1$ is a hypergeometric function, and $\gamma \equiv \frac{\alpha^2}{\alpha^2 +1}, \eta \equiv \frac{q^2}{\beta^2 r^4} \left(\frac{r}{b}\right)^{4\gamma}$. Parameters $m$ and $q$ are related to the mass $M$ and charge $Q$ of the BH respectively,
\begin{eqnarray}
M = \frac{m b^{2 \gamma}}{2 (\alpha^2 +1)},\qquad Q = q.
\end{eqnarray}
From the metric, the Hawking temperature and Bekenstein-Hawking entropy of the BH can be obtained
\begin{eqnarray}\label{Hawking-temperature}
T &=& \frac{\alpha^2 +1}{2\pi r_+ (\alpha^2 - 1)} \left(\frac{b}{r_+}\right)^{2\gamma} \left[r_+^2 (\alpha^2 - 1) \left(\beta^2 -\frac{\Lambda}{2}\right) - \frac{1}{2} \left(\frac{r_+}{b}\right)^{4\gamma} - \beta^2 r_+^2 (\alpha^2 - 1) \sqrt{\eta_+ + 1} \right],\nonumber\\
S &=& \pi r_+^2 \left(\frac{b}{r_+}\right)^{2\gamma}.
\end{eqnarray}
with $\eta_+ \equiv \eta\big|_{r=r_+}$ and $r_+$ is the outmost horizon radius.

We will study the phase transitions in the extended phase space where the negative cosmological constant is treated as a positive thermodynamical pressure. In our case, the precise relation is
\begin{eqnarray}
P = -\frac{\Lambda}{8\pi} \left(\frac{b}{r_+}\right)^{2\gamma},
\end{eqnarray}
which reduces to the standard relation in the limit of $\alpha=0$. With the help of Eq.~(\ref{Hawking-temperature}), the above equation can be cast into an equation of state
\begin{eqnarray}
P = P(r_+, T) = \frac{T}{2 r_+ (\alpha^2 +1)} + \frac{1}{8\pi r_+^2} \left[\frac{1}{\alpha^2 -1} \left(\frac{b}{r_+}\right)^{-2\gamma} - 2 \beta^2 r_+^2 \left(\frac{b}{r_+}\right)^{2\gamma} \left(1 - \sqrt{\eta_+ + 1}\right)\right].
\end{eqnarray}
We will work in the canonical ensemble of the extended phase space where $Q$ and $\beta$ is fixed. And $\alpha$ and $b$ will be considered as constants since they do not enter into the first law of thermodynamics. The phase transition is governed by the Gibbs free energy, which is
\begin{eqnarray}
G &=& M - T S =\frac{1}{4}\Bigg\{2\beta^2 r_+^3 (\alpha^2 + 1) \left(\frac{b}{r_+}\right)^{4\gamma} \left(\sqrt{\eta_+ + 1} -1 \right)  - r_+ \left[8\pi r_+^2 P (\alpha^2 + 1) \left(\frac{b}{r_+}\right)^{2\gamma} - 1\right] \nonumber\\
&&- \frac{4(\alpha^2 + 1)}{\alpha^2 -3} b^{4\gamma} r_+^{(3-\alpha^2)/(\alpha^2 + 1)} \left[4 \pi P \left(\frac{b}{r_+}\right)^{-2\gamma} + \beta^2 \left(1-\ _2{\cal F}_1 \left(-\frac{1}{2}, \frac{\alpha^2 -3}{4}, \frac{\alpha^2 +1}{4}, -\eta_+\right) \right)\right]  \Bigg\}.\nonumber\\
\end{eqnarray}
By studying the behavior of the Gibbs free energy versus the temperature for fixed pressure, the phase structures and transitions of the system can be depicted. The full $P-T$ phase diagram is rather rich and its exact picture depends on the set of free parameters $(q, \alpha, \beta, b)$.

\subsection{Born-Infeld-dilaton black holes (finite $\beta$)}

In the following discussions, we will fix $q=0.2, b=1$ and $\beta=2$ to study the effect of $\alpha$ on the phase diagram. We should note that $\alpha <1$ to ensure the positivity of the Hawking temperature. As $\alpha$ is increased starting from zero, four types of phase diagrams will appear consequently for each we will only show one example in the following.

\begin{itemize}
\item {\bf Case (a):} $\alpha=0.008$

\begin{figure}[!htbp]
\includegraphics[width=0.45\textwidth]{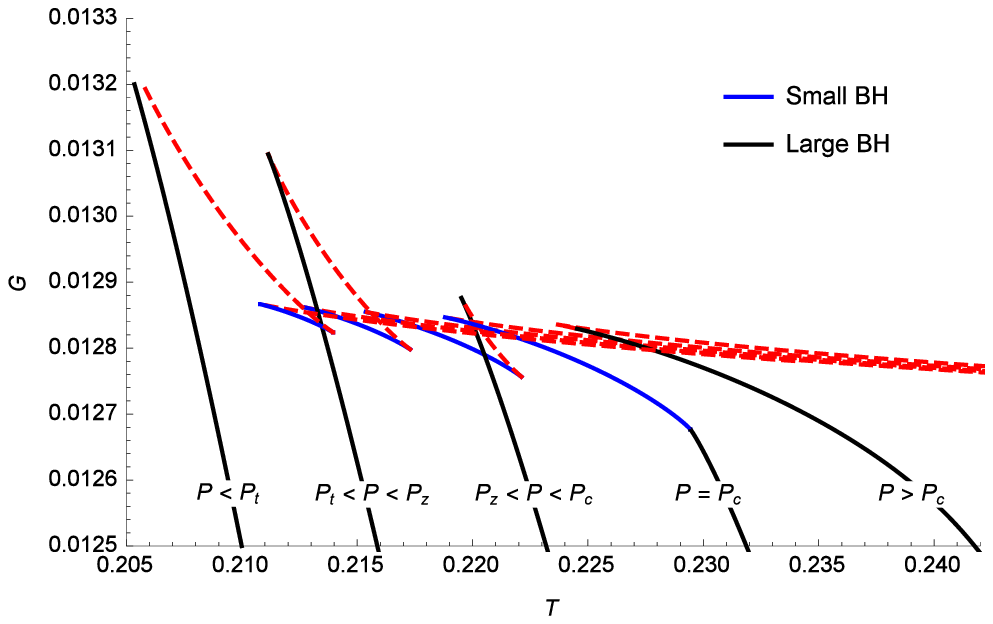}\quad
\includegraphics[width=0.45\textwidth]{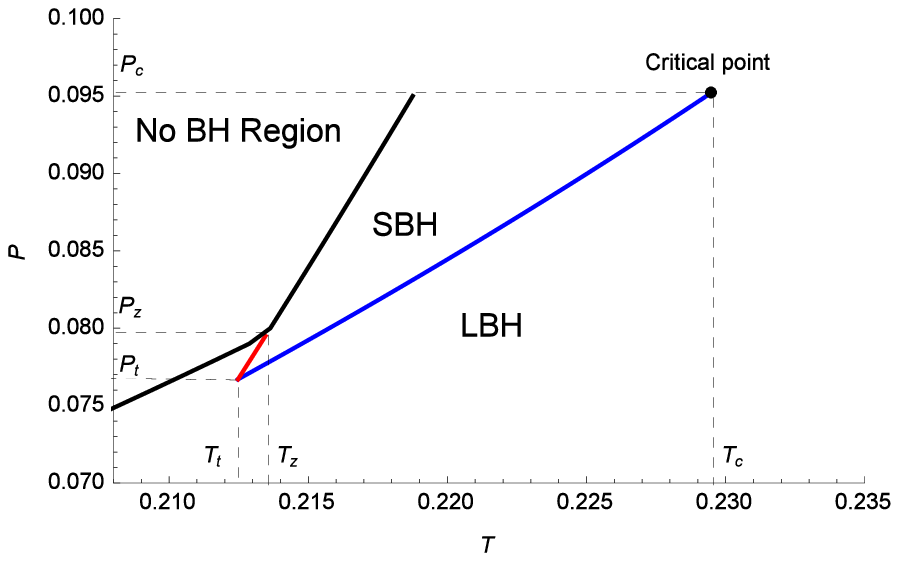}
\caption{(colour online) The Gibbs free energy $G$ as a function of $T$ for fixed pressure (Left) and $P-T$ diagram diagrams (Right) in case (a). Red dashed lines in the left panel correspond to unstable branches of BH solutions. $(T_c, P_c)$ is the critical point. In the $P-T$ diagram, the blue solid line denotes the co-existing line of first-order LBH-SBH phase transition, while the red solid line denotes the co-existing line of zeroth-order SBH-LBH phase transition.}
\end{figure}

In this case, we have the common $P-T$ phase diagram containing RPT of BHs first observed in Refs.~\cite{Gunasekaran:2012dq,Altamirano:2013ane}, as shown in Fig.~1. When $P_z<P<P_c$, there is a first order LBH-SBH phase transition; When $P_t<P<P_z$, a LBH-SBH-LBH RPT appears with LBH-SBH transition being first-order and SBH-LBH zeroth-order; When $P<P_t$, no phase transition occurs.

\item {\bf Case (b):} $\alpha=0.1$

\begin{figure}[!htbp]
\includegraphics[width=0.45\textwidth]{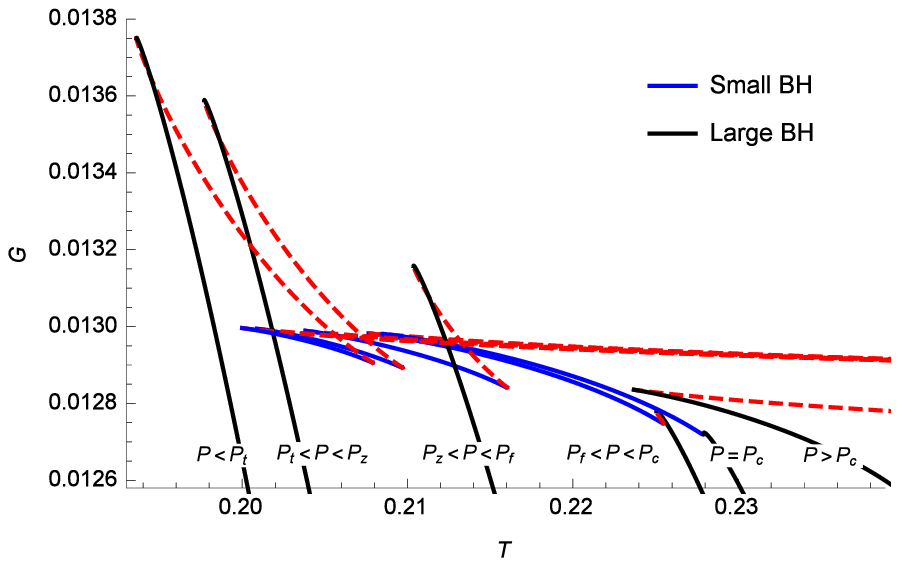}\quad
\includegraphics[width=0.45\textwidth]{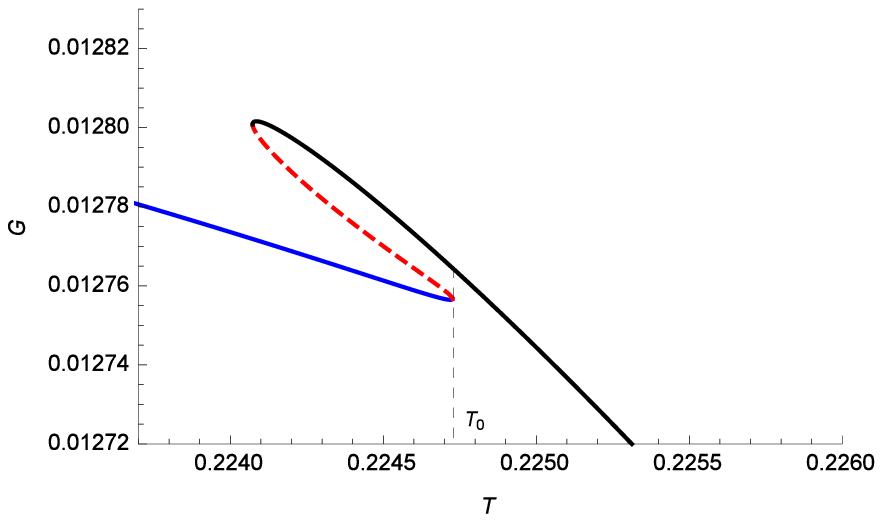}
\includegraphics[width=0.45\textwidth]{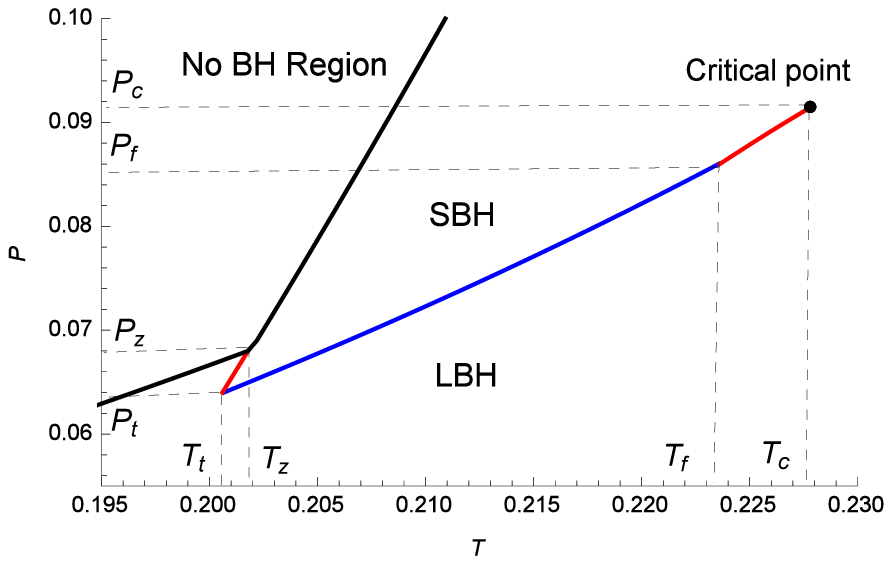}
\caption{(colour online) $G-T$ and $P-T$ diagrams in case (b). In the first line, the right panel is a close-up view of the curve with $P_f<P<P_c$ in the left panel, which shows a zeroth-order LBH-SBH phase transition with transition temperature $T_0$. In the $P-T$ diagram in the second line, the additional red solid line in the temperature range $[T_f,T_c]$ denotes the co-existing line of the zeroth-order LBH-SBH phase transition.}
\end{figure}

As shown in Fig.~2, in this case, compared to case (a), there is an additional zeroth-order LBH-SBH transition for $P_f<P<P_c (P_f > P_z)$ where the Gibbs free energy experiences a discontinuity at the transition temperature $T_0$.

\item {\bf Case (c):} $\alpha=0.6$

\begin{figure}[!htbp]
\includegraphics[width=0.45\textwidth]{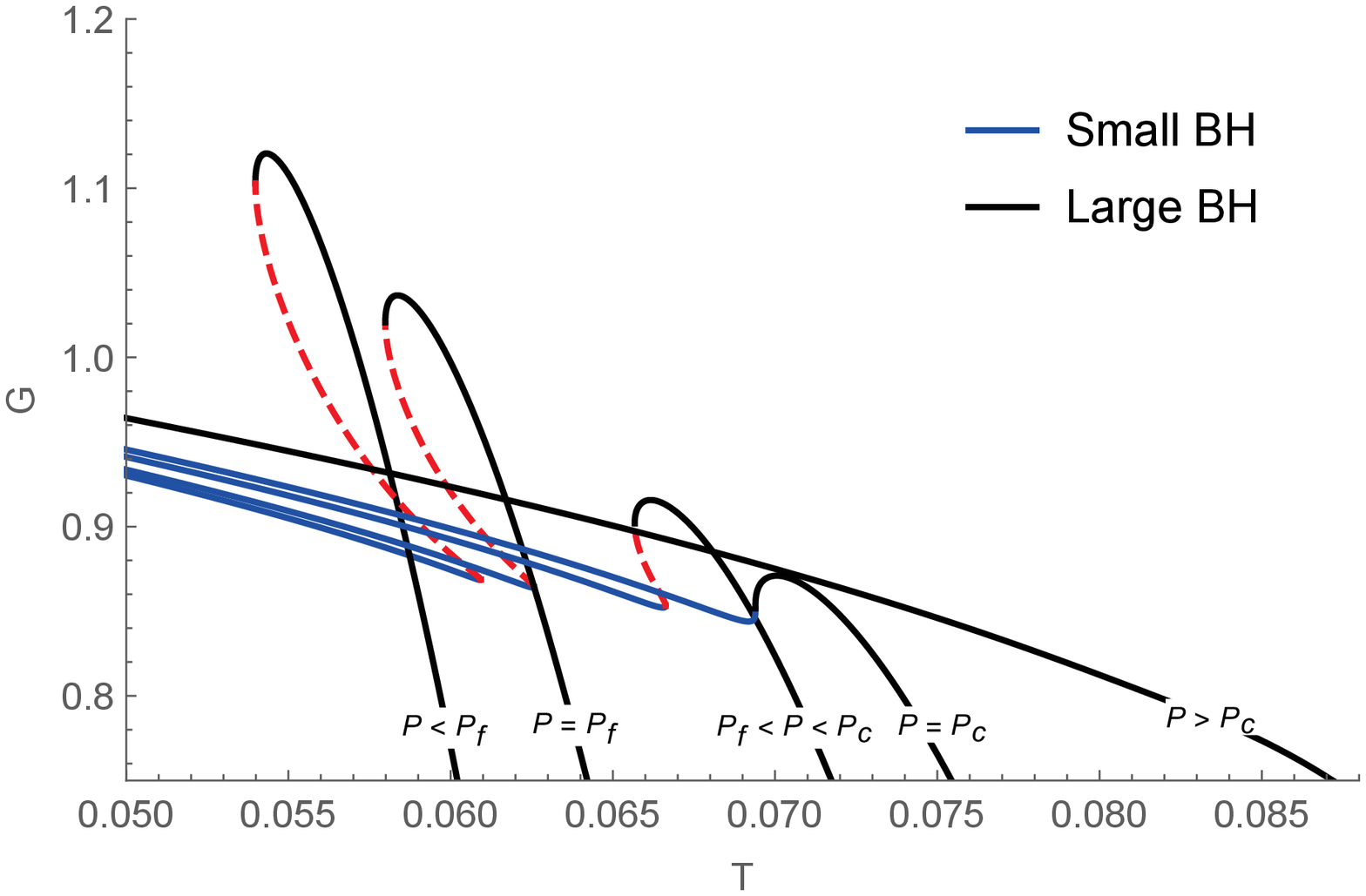}
\includegraphics[width=0.45\textwidth]{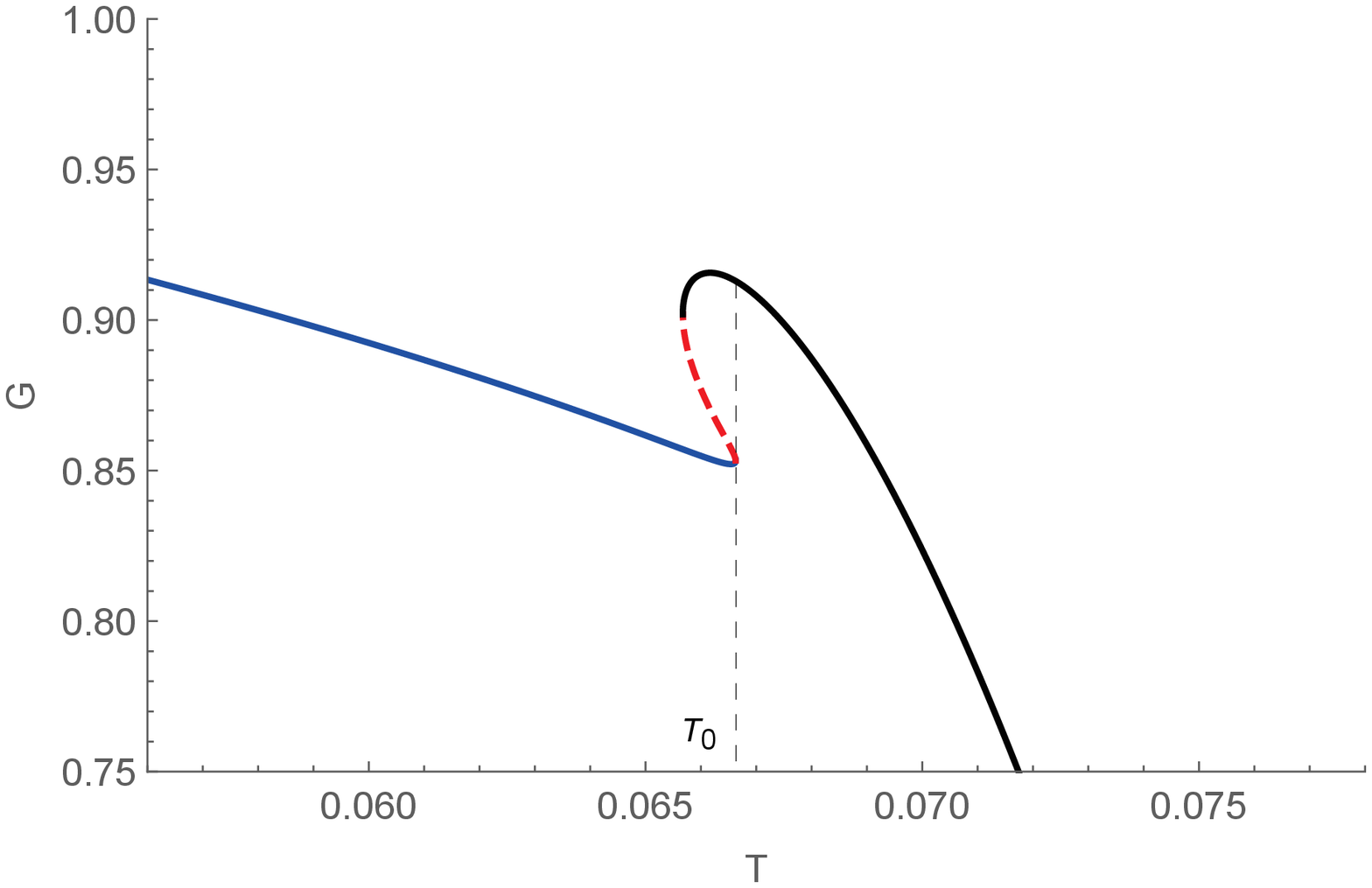}
\includegraphics[width=0.45\textwidth]{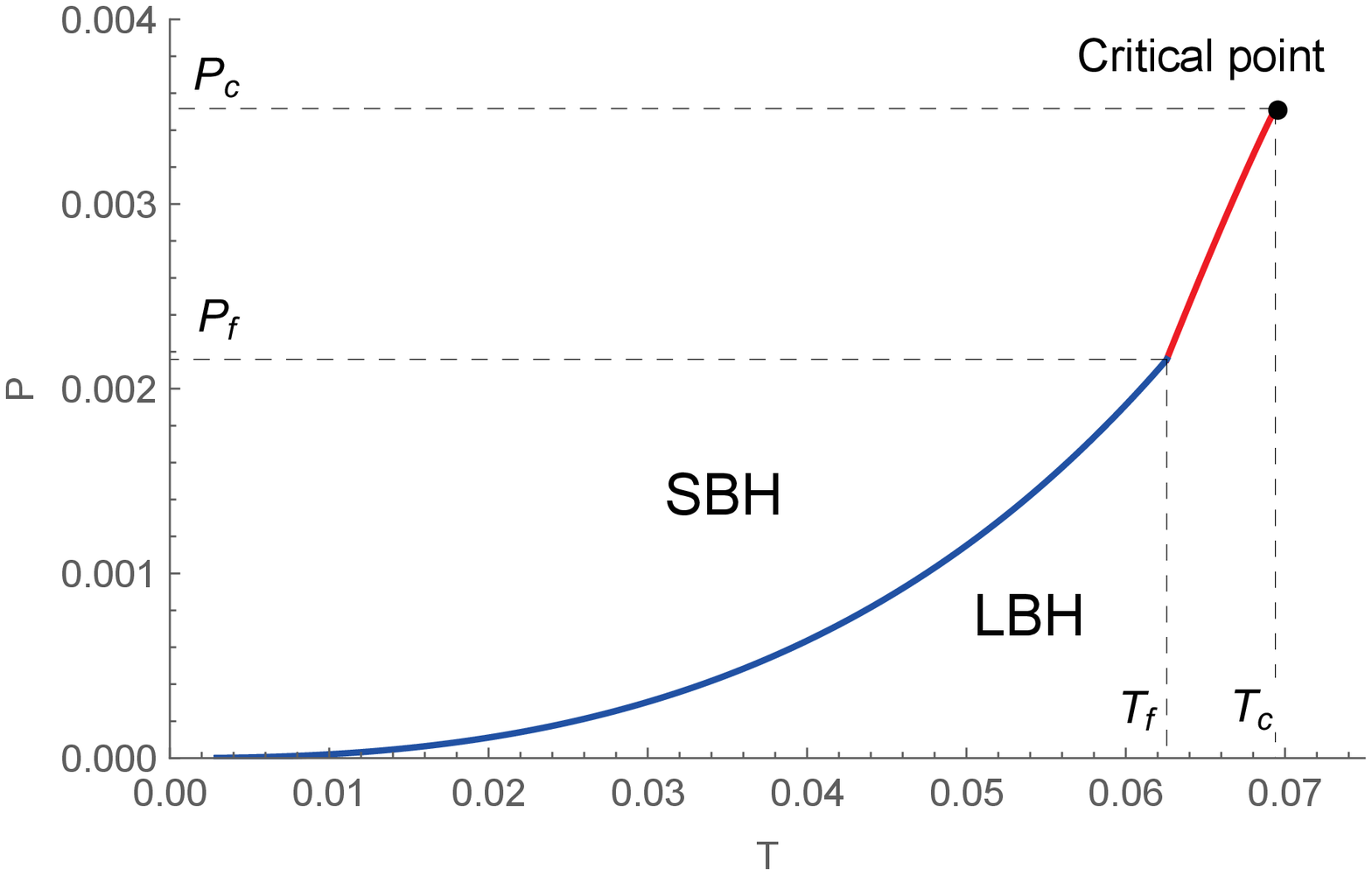}
\caption{(colour online) $G-T$ and $P-T$ diagrams in case (c). In the first line, the right panel is a close-up view of the curve with $P_f<P<P_c$ in the left panel, which shows a zeroth-order LBH-SBH phase transition with transition temperature $T_0$.}
\end{figure}

\begin{figure}[!htbp]
\includegraphics[width=0.45\textwidth]{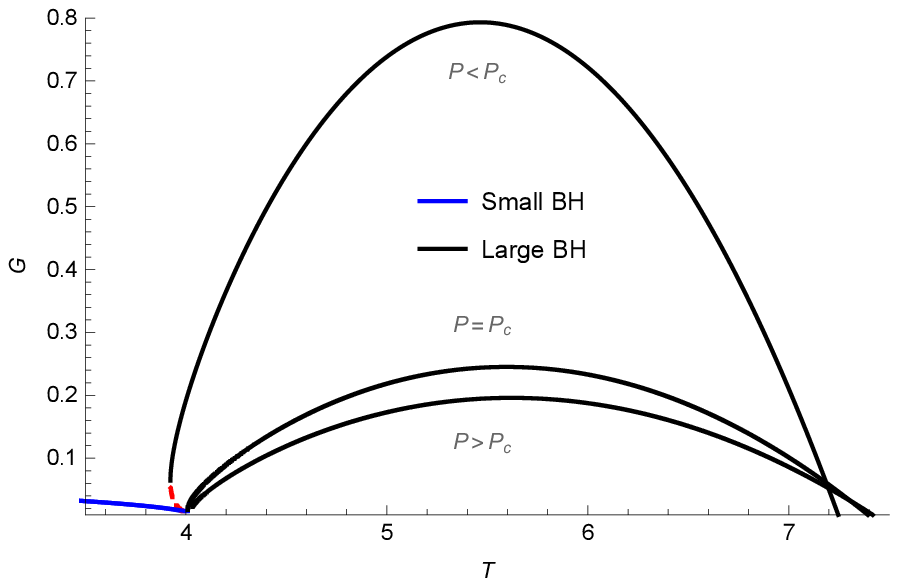}
\includegraphics[width=0.45\textwidth]{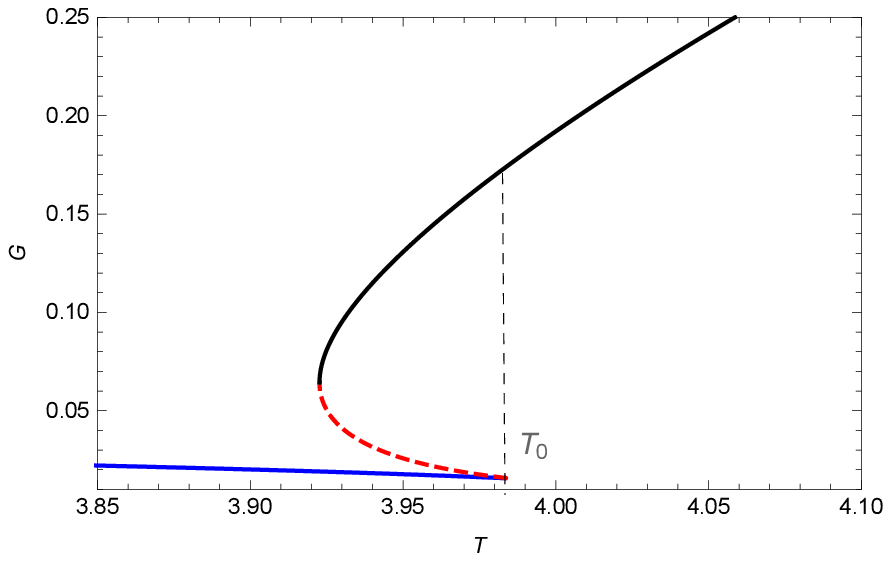}
\includegraphics[width=0.45\textwidth]{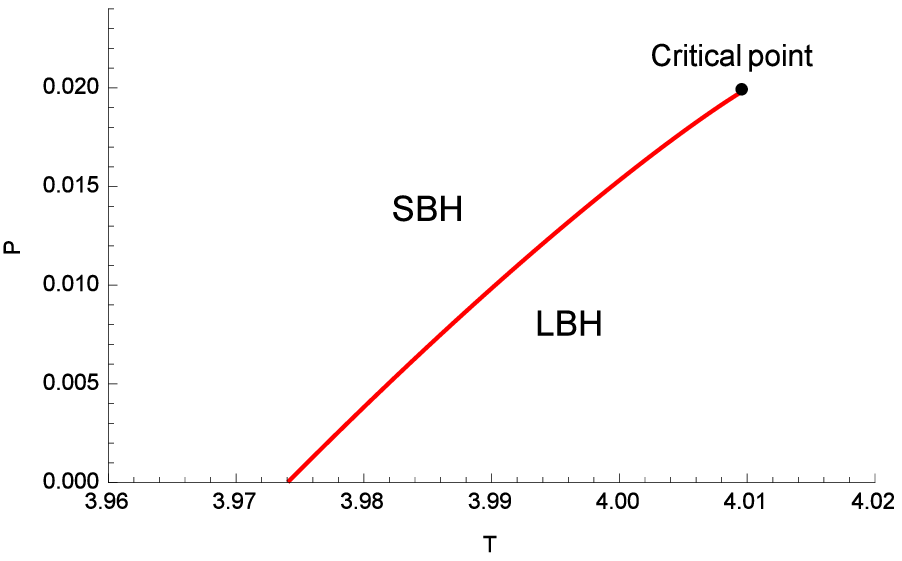}
\caption{(colour online) $G-T$ and $P-T$ diagrams in case (d). In the first line, the right panel is a close-up view of the curve with $P<P_c$ in the left panel, which shows a zeroth-order LBH-SBH phase transition with transition temperature $T_0$.}
\end{figure}

In this case, compared to case (b), the RPT region disappears and there remain only zeroth-order and first-order phase transitions between LBH and SBH: When $P_f<P<P_c$, it is zeroth-order while for $P<P_f$ it is first-order, as shown in Fig.~3.

\item {\bf Case (d):} $\alpha=0.98$

In this case, compared to case (c), the first-order phase transition region disappears and only the zeroth-order phase transition remains as shown in Fig.~4.
\end{itemize}

\subsection{Maxwell-dilaton black holes ($\beta \rightarrow \infty$)}

The Maxwell-dilaton BH can be obtained by taking the limit of $\beta \rightarrow \infty$ of the above mentioned Born-Infeld-dilaton BH. The BH contains three parameters $(q,b,\alpha)$. In the following, we fix $q=1$ and $b=1$ to study the effect of $\alpha$ on the phase diagram. As $\alpha$ is increased from zero to the upper bound, two types of phase diagrams will appear consequently.

\begin{itemize}
\item {\bf Case (e):} $\alpha=0$

\begin{figure}[!htbp]
\includegraphics[width=0.45\textwidth]{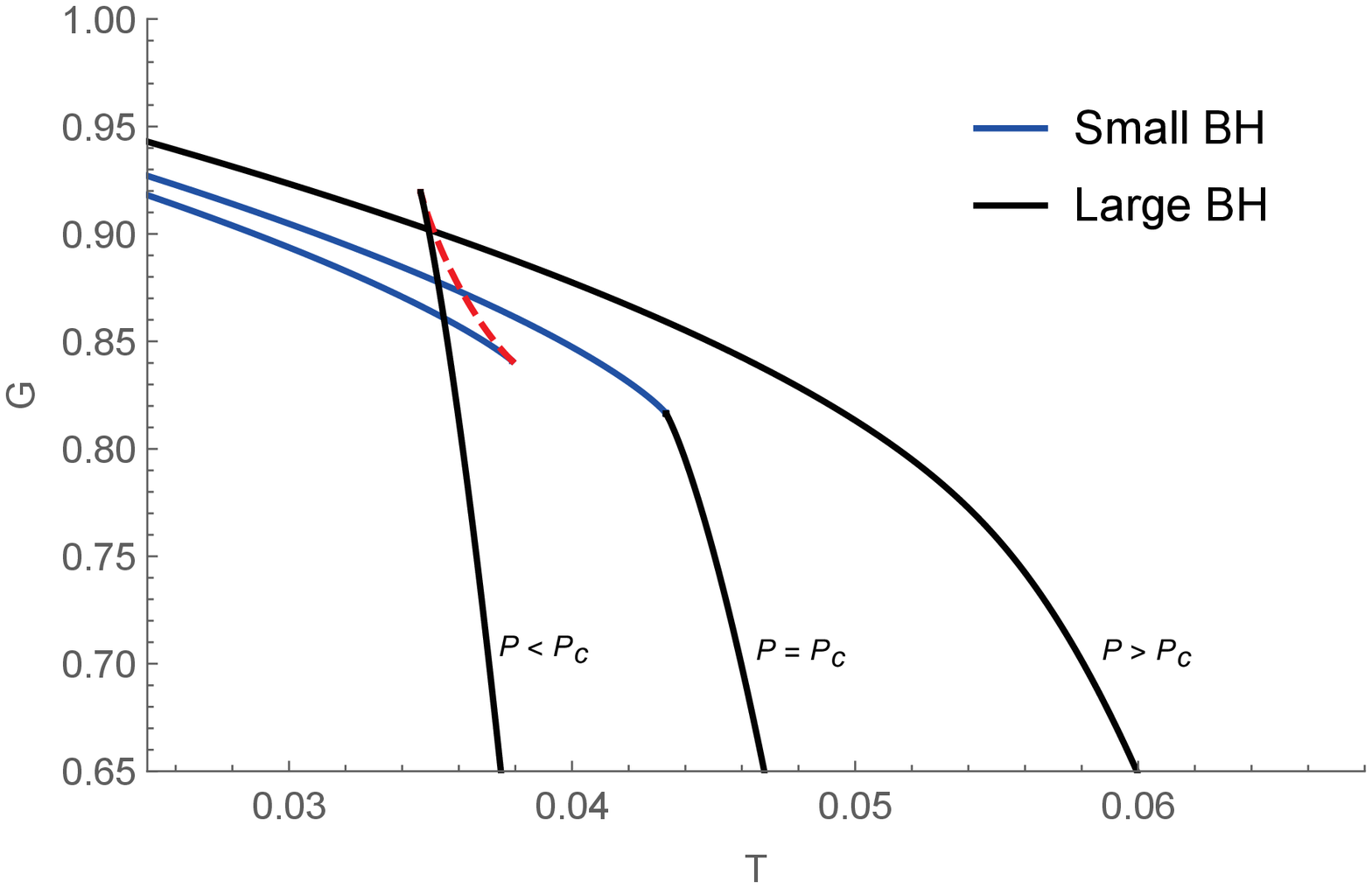}
\includegraphics[width=0.45\textwidth]{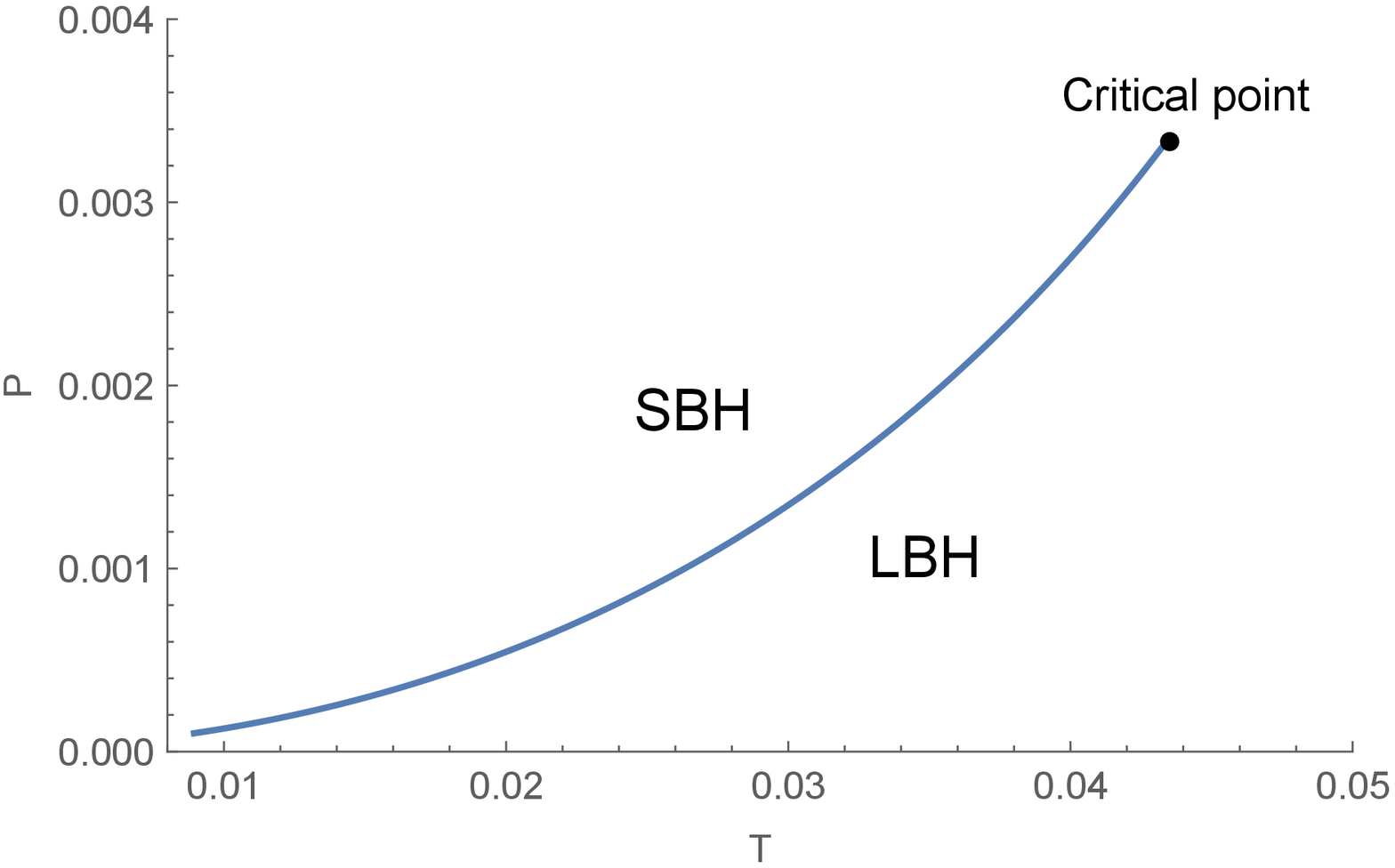}
\caption{(colour online) $G-T$ and $P-T$ diagrams in case (e).}
\end{figure}

In this case, as shown in Fig.~5, the $P-T$ phase diagram resembles that of a Van de Waals system: there is only a first-order LBH-SBH phase transition for $P<P_c$. No RPTs exist.

\item {\bf Case (f):} $\alpha>0$

This case is the same as case (c). Compared to case (e), an additional zeroth-order LBH-SBH
phase transition appears for $P_f< P < P_c$.
\end{itemize}

\section{Ruppeiner geometry}

In this section, we will study the above mentioned phase transitions by adopting the Ruppeiner geometry. We define the Ruppeiner metric in $X^a = (M, P)$ space as
\begin{eqnarray}
g_{ab} = - \frac{\partial^2 S}{\partial X^a \partial X^b}.
\end{eqnarray}
In practical calculations~\footnote{We should note that the metric defined here is not guaranteed to be positive definite. Actually, the determinant of the metric in our cases is always negative, thus the standard thermodynamical stability criterions, as Eqs.(22-24) in Ref.~\cite{Ruppeiner:2013yca}, can not be satisfied. So we should be caution to use Eq.~(25) in Ref.~\cite{Ruppeiner:2013yca} to calculate $R$.}, it is more convenient to cast the above form into the Weinhold form
\begin{eqnarray}
g_{ab} = \frac{1}{T} \frac{\partial^2 M}{\partial Y^a \partial Y^b},
\end{eqnarray}
where $Y^a = (S, P)$. We should note that our definition of the Ruppeiner metric is different from the standard one where the metric is defined instead on the $(M, Q)$-space. The corresponding Ricci scalar $R$, called Ruppeiner invariant, can give us some insights into the microscopic structures and behaviors of possible BH molecules living on the BH horizon. It is conjectured that the amplitude $|R|$ measures the average number of correlated constituents (and thus measures the strength of the interaction between constituents), and its sign indicates the dominant interaction between the BH molecules: $R>0, R<0$ and $R=0$ indicate repulsion, attraction and no interaction respectively.

\begin{figure}[!htbp]
\centering
\subfigure[~~The first type with RPT]{\includegraphics[width=0.45\textwidth]{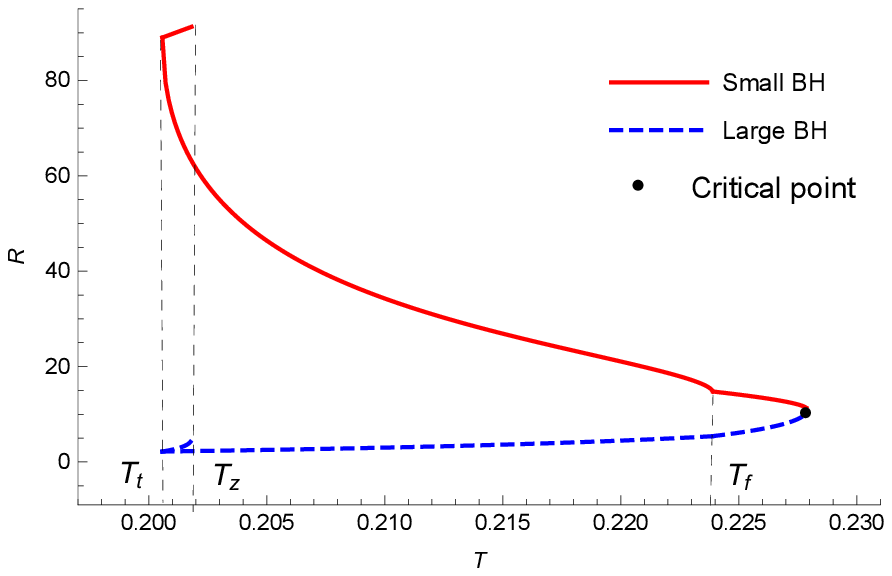}}\quad
\subfigure[~~The second type without RPT]{\includegraphics[width=0.45\textwidth]{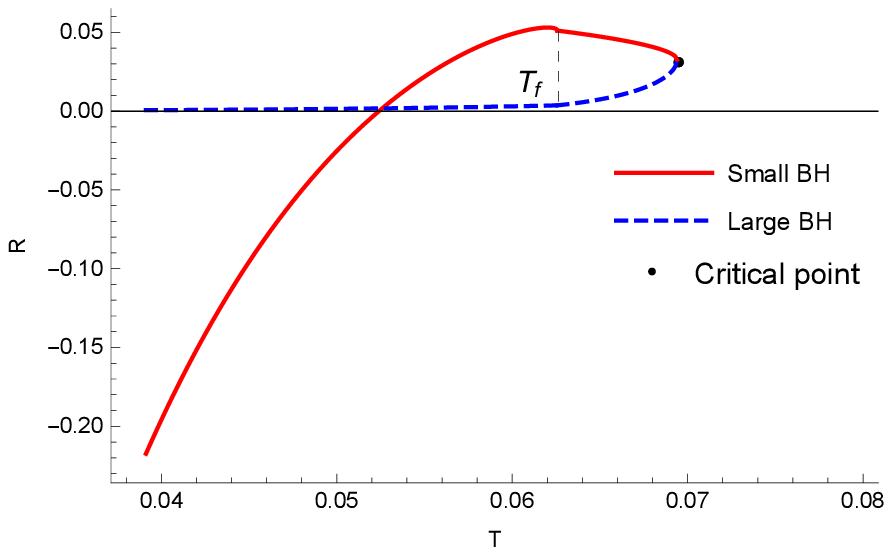}}
\caption{(colour online) Two types of $R-T$ pictures along the co-existing lines. Actually, the left panel is for case (b) and the right panel is for case (f). For case (a), there is no zeroth-order phase transition, so the $[T_f,T_c]$-segment in the left panel will shrink to the critical point. For case (c), the picture is similar to case (f); For case (d), there is no zeroth-order phase transition, so the segment with $T<T_f$ in the right panel will disappear; And for case (e), there is no first-order phase transition, so the $[T_f,T_c]$-segment in the right panel will shrink to the critical point.}
\end{figure}

With the definition of the Ruppeiner metric, we calculate the Ruppeiner invariant $R$ along the co-existing lines for all the cases (a-f). We observe that the behaviors of $R$ versus the transition temperature along the co-existing lines can be classified into two types according to whether there exists RPT region or not. The first kind is for cases (a)(b) with RPT region and the second one is for cases (c-f) without RPT region. For simplicity, in Fig.~7, we only show the $R-T$ pictures along the co-existing lines for case (b) and case (f) as they have the richest phase diagram in each kind. Other cases will have similar pictures with the corresponding segment of the $R-T$ lines disappearing if the first-order/zeroth-order phase transition is absent.

From the figure, we can see that the main difference between the two kinds of $R-T$ pictures along the co-existing lines lies in the behaviors of $R$ for SBH. In the first type with RPT shown in the left panel, $R$ for SBH is always positive and increases quickly and monotonically as the transition temperature is lowered, and will become very large in the RPT range $T \in [T_t, T_z]$; While in the second type without RPT shown in the right panel, as the transition temperature is lowered, $R$ for SBH experiences two stages: it increases firstly to reach a maximum (positive but small) value and then decreases quickly to be negative. For LBH, $R$ has similar behavior for the two types and is always positive but close to zero. We should note the the multi-values of $R$ in the RPT range $T \in [T_t, T_z]$ in the left panel are due to the successive two transitions (LBH-SBH-LBH) of RPT.

With these observations, we may explain the various of phase transitions, especially the RPT, as a result of two competing factors: The first one is the low-temperature effect which tends to freeze the BH molecules and thus shrink the BH, while the second one is the repulsive interactions between BH molecules which, on the contrary, tends to expand the BH. From the left panel of the figure, we can see that outside the RPT range $T \in [T_z, T_c]$, $R$ for LBH (and also for SBH) is positive but small which means weakly repulsive interaction between BH molecules, and thus the first kind of factor dominates over the second one and LBH tends to shrink and thus transits to SBH; While in the RPT range $T \in [T_t, T_z]$, after the LBH-SBH transition, $R$ for SBH becomes very large which means strong repulsion between BH molecules and it will become even larger as temperature is further lowered, so finally the repulsion may become strong enough to dominate over the first factor and then SBH tends to expand to release the high repulsion and transits back to LBH. From the right panel of the figure, we can see that in this type $R$ for SBH has a maximum positive value and even becomes negative when the temperature is further lowered, so the first factor is always dominate over the second one and after the LBH-SBH transition there is no chance for the SBH to transit back to LBH, so no RPT occurs.

\begin{figure}[!htbp]
\centering
\subfigure[~~The first type with RPTs]{\includegraphics[width=0.45\textwidth]{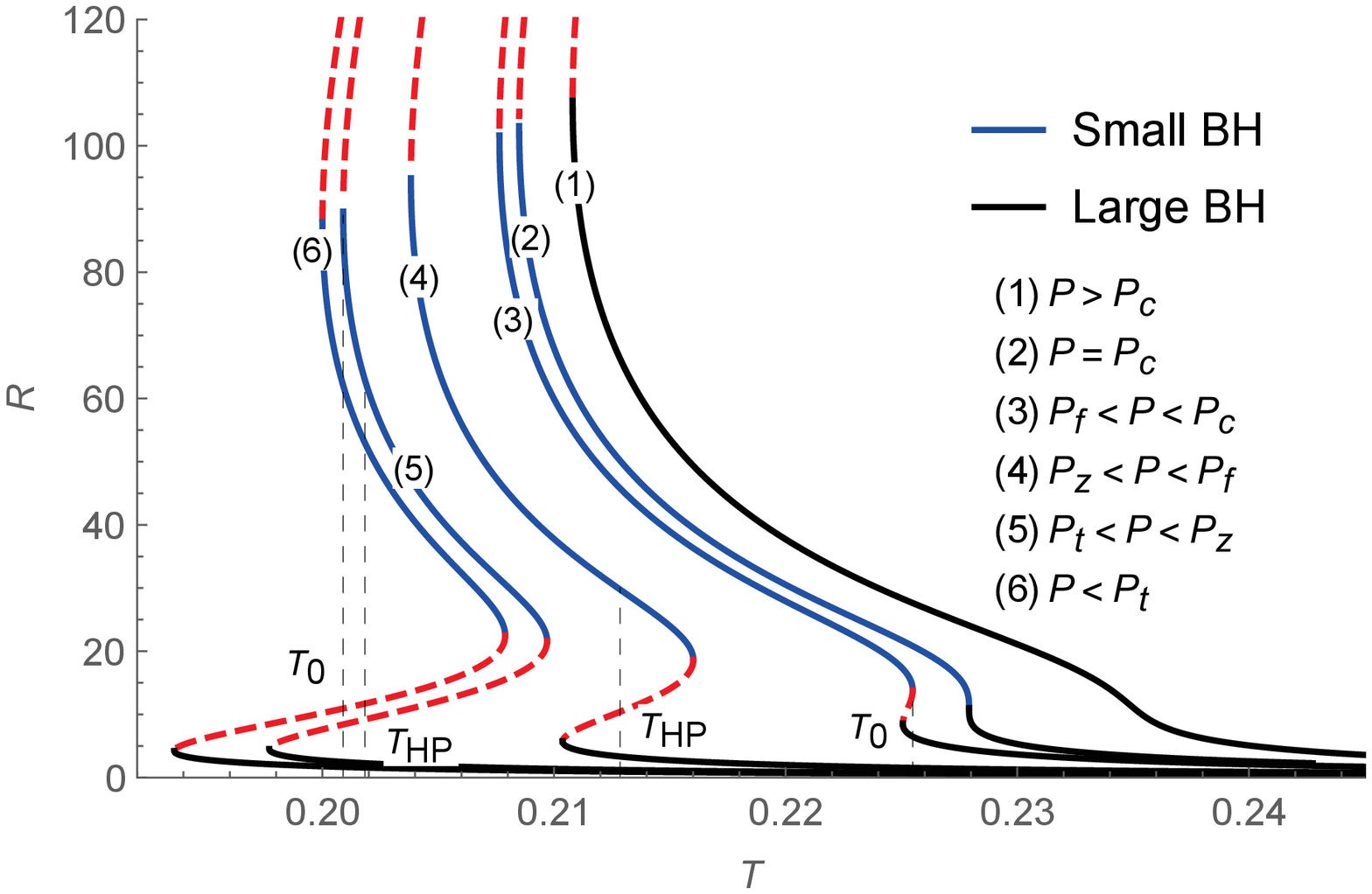}}
\subfigure[~~The second type without RPTs]{\includegraphics[width=0.45\textwidth]{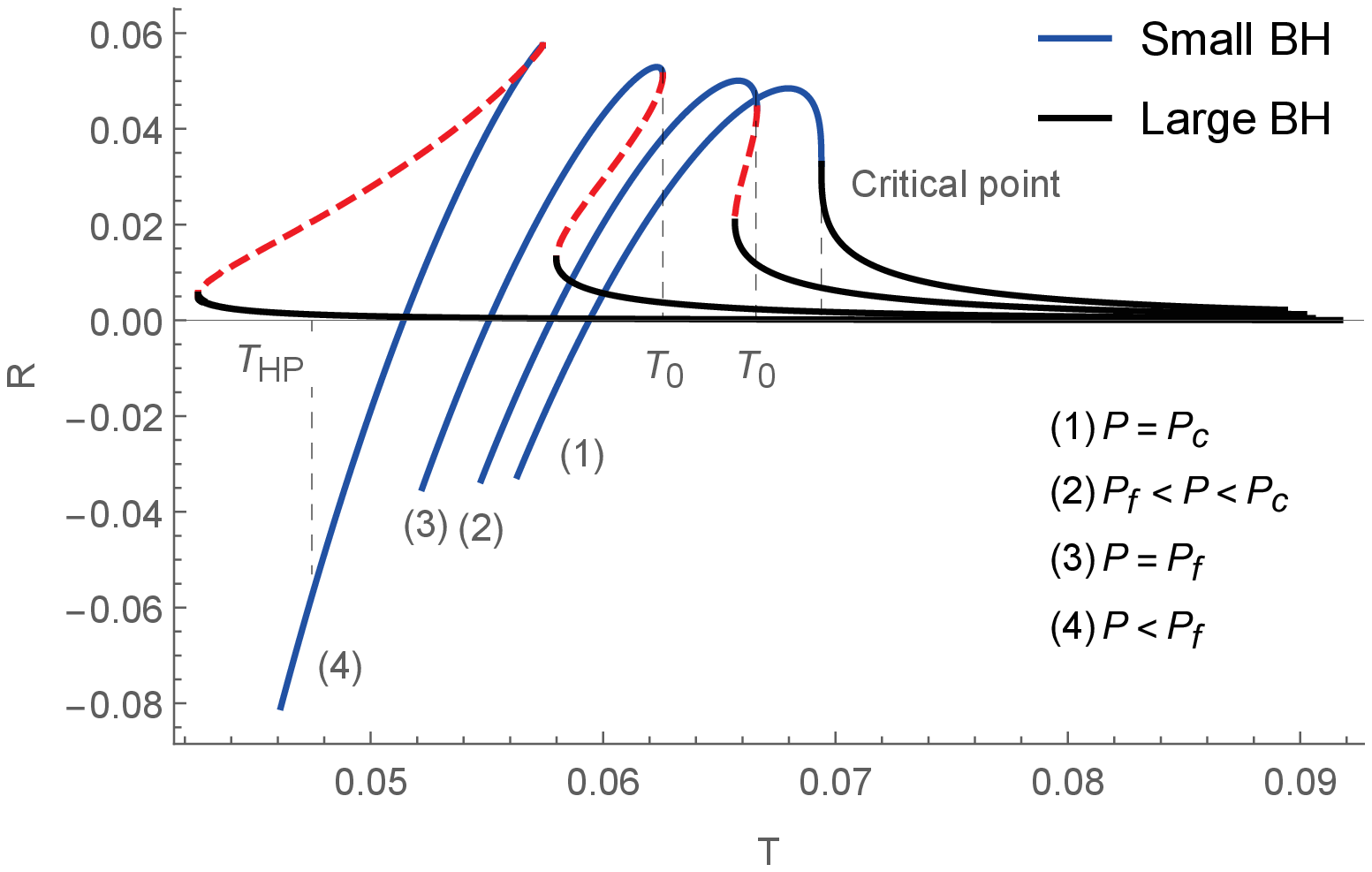}}
\caption{(colour online) Two Types of $R-T$ pictures with fixed pressure $P$. $T_0$ and $T_{\rm HP}$ respectively denote the transition temperatures for zeroth-order and first-order phase transitions. Dashed red lines denote unstable branches of BH solutions.}
\end{figure}

We can also gain this explanation from Fig.~7, in which we plot $R$ as a function of the temperature $T$ for fixed pressure $P$. From the figure, we can see that in the first type with RPT, $R$ for SBH increases monotonically and quickly as $T$ is lowered; While in the second type without RPT, $R$ for SBH has a maximum (positive but small) value at certain temperature and it then decreases quickly to be negative as $T$ is further lowered. For LBH, $R$ has similar behavior for the two types and is always positive but close to zero, which looks like a ideal gas system. Compared to ordinary two-dimensional thermodynamical systems but with fixed specific volume~\cite{Ruppeiner:2013yca}, it is interesting to see that SBH behaves like a Fermionic gas system in cases with RPT, while it behaves (or partly) oppositely to an anyon system in cases without RPT. And in all cases, LBH behaves like a nearly ideal gas system.

\section{Summary and discussions}

In this work, we try to give a microscopic explanation of the various phase transitions of charged dilatonic BHs via Ruppeiner geometry. Our results show that the various phase transition, including zeroth/first-order LBH-SBH transition and RPT, may be uniformly well explained as a result of two competing factors. One is the low-temperature effect which tends to shrink the BH and the other one is the repulsive interaction between BH molecules which, on the contrary, tends to expand the BH. Which factor dominates depends on the value of the coupling constant $\alpha$, the pressure $P$ and the temperature $T$. In cases without RPT, as $T$ is decreased, the low-temperature effect dominate over the repulsion between BH molecules, so that LBH tends to shrink and thus transits to SBH. While in the cases with RPTs, after the LBH-SBH transition, as $T$ is further lowered, the repulsion effect increases quickly (as we can see from Figs.~6 and 7) and finally becomes strong enough to dominate over the low-temperature effect, so that SBH tends to expand to release the high repulsion and thus transits back to LBH.

By comparing the behavior of $R$ versus $T$ for fixed pressure to that of ordinary two-dimensional thermodynamical systems but with fixed specific volume~\cite{Ruppeiner:2013yca}, it is interesting to see that SBH behaves like a Fermionic gas system in cases with RPT, while it behaves (or partly) oppositely to an anyon system in cases without RPT. And in all cases, LBH behaves like a nearly ideal gas system.

In this paper, we only consider charged dilatonic BHs in Einstein-Born-Infeld-dilaton or Einstein-Maxwell-dilaton theories. It is interesting to see whether this picture is applicable for other BHs in other gravity theories or other types of phase transitions. We leave it for further investigations.

\section*{Acknowledgement}

This work was supported by National Natural Science Foundation of China (Nos. 11605155 and 11675144).

\end{document}